\documentclass[openacc]{rstransa}

\usepackage{subfig}
\usepackage{float}
\usepackage{comment}

\begin{document}

\title{Observing a Dynamical Skeleton of Turbulence in Taylor-Couette Flow Experiments}

\author{
C. J. Crowley$^{1}$, J. L. Pughe-Sanford$^{1}$, W. Toler$^{1}$, R. O. Grigoriev$^{1}$, and M. F. Schatz$^{1}$}

\address{$^{1}$Center for Nonlinear Science and School of Physics, Georgia Institute of Technology, 837 State Street, Atlanta, Georgia 30332, USA}

\subject{Fluid Mechanics, Dynamical Systems, Turbulence, Spatio-temporal Chaos}


\corres{Michael F. Schatz\\
\email{michael.schatz@physics.gatech.edu}}

\begin{abstract}
Recent work suggests unstable recurrent solutions of the equations governing fluid flow can play an important role in structuring the dynamics of turbulence. Here we present a method for detecting intervals of time where turbulence "shadows" (spatially and temporally mimics) recurrent solutions. We find that shadowing occurs frequently and repeatedly in both numerical and experimental observations of counter-rotating Taylor-Couette flow, despite the relatively small number of known recurrent solutions in this system. Our results set the stage for experimentally-grounded dynamical descriptions of turbulence in a variety of wall-bounded shear flows, enabling applications to forecasting and control.
\end{abstract}


\begin{fmtext}

\section{Introduction}

At the start of his 1923 tour de force \cite{Taylor1923}, G. I  Taylor stated, "A great many attempts have been made to discover some mathematical representation of fluid instability, but so far they have been unsuccessful in every case."  Taylor illustrated, for the first time, how appropriate solutions to the equations governing fluid motion can be harnessed to make predictions about unstable behaviors that may be tested quantitatively in experiments. Specifically, in the case of flow between concentric, independently-rotating cylinders (Taylor-Couette Flow (TCF)), Taylor demonstrated how unstable solutions to the governing equations, linearized about a base state (Couette flow), can signal the transition to a new stable, non-turbulent flow (e.g. Taylor Vortex Flow); the remarkable quantitative agreement between predictions and observations that Taylor found

\end{fmtext}


\maketitle

\noindent established linear stability analysis as a core methodology for the prediction of flow transitions in TCF, Rayleigh-B\'enard convection and numerous other
problems in fluid mechanics \cite{cross1993}. Following in Taylor's footsteps, we study the same system (TCF) and also show how knowledge of unstable solutions can be exploited to make experimentally-testable predictions; however, in the present work, the behaviors of unstable solutions themselves, in their full non-linear glory, are of primary interest, rather than those of linearized versions, which serve chiefly as indicators of shifts between  stable nonlinear states.  

Some twenty years after the publication of Taylor's work, Eberhard Hopf speculated the dynamics of unstable solutions are central to understanding turbulence \cite{Hopf1942, Hopf1948}.  In essence, Hopf envisioned that turbulence results from sequential visits to members of a repertoire of unstable solutions. Long-established observations of turbulent coherent structures--- fleeting flow patterns that recur repeatedly \cite{Hussain1983,Jimenez2018}---suggest a connection to unstable solutions.  The advent of advanced numerical methods such as Newton-Krylov solvers \cite{Viswanath2007} have enabled the computation of unstable, recurrent solutions, which can  exhibit features qualitatively similar to those of coherent structures observed in the laboratory \cite{itano2009}. Consequently, such unstable solutions have become known as exact coherent structures (ECSs). 
 
 We show ECSs play important and persistent roles in guiding the time-evolution of three-dimensional, experimentally observable turbulence.
 The simplest type of ECSs that occur in turbulence are equilibria \cite{Suri2017, Suri2018} and, in systems possessing a continuous spatial symmetry (e.g. circular symmetry in TCF), traveling waves \cite{Faisst2003, Hof2004, deLozar2012, Lemoult2014, Park2015, Mellibovsky2009, kerswell2007,Eckhardt2018b}, which are equilibria in suitable co-moving reference frames.  ECSs in the form of periodic orbits \cite{Cvitanovic2010,Kawahara2001} also arise and have been found to characterize turbulent behaviors in quasi-two-dimensional flow experiments \cite{Suri2020}.  In systems with a continuous symmetry, snapshots of turbulence in numerical simulations \cite{Budanur2017} are found to resemble unstable relative periodic orbits (RPOs), which are periodic orbits in a co-moving reference frame; moreover, RPOs can capture the time-evolution of turbulence in numerics \cite{Krygier2021}.   Here we provide convincing evidence that TCF turbulence "shadows" RPOs---that is, turbulence frequently and repeatedly emulates the spatial {\it and} temporal structure of one or more ECS in the form of RPOs.

The paper is organized as follows: The particulars of TCF in this study are outlined in Section \ref{sec:es}, and key aspects of the ECSs identified and explored herein are described in Section \ref{sec:ecstcf}.  The phenomenon of shadowing is described from a state-space perspective in Section \ref{sec:ssg} and the procedures for detecting shadowing in turbulence experiments are discussed in Section \ref{sec:detect}.  Results are presented in Section \ref{sec:results} and conclusions are found in Section \ref{sec:conc}.

\begin{figure}[]
    \center
    \includegraphics[width=.65\textwidth]{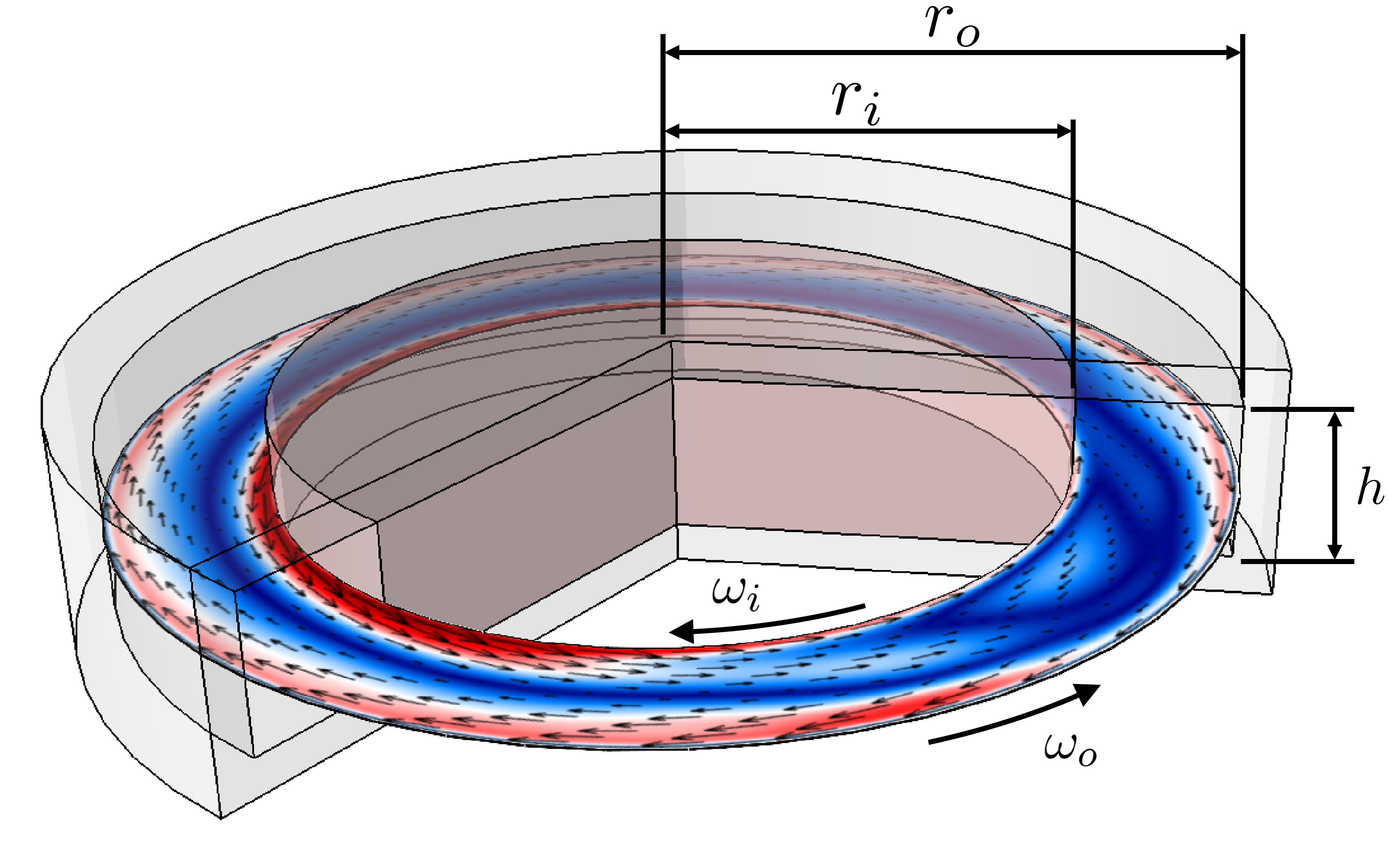} 
    \caption{\label{fig:TCF_geometry} Turbulence is visualized in a laboratory flow between concentric, independently-rotating cylinders with radii $r_i$, $r_o$ and corresponding angular velocities $\omega_i,\omega_o$. Fluid is confined between the cylinders and bounded axially by end caps co-rotating with the outer cylinder. 
    The azimuthal velocity, $u_\theta$, is plotted in the horizontal cross section, $z=1.5$\,mm, of the flow volume as a black vector field. The colormap indicates the magnitude of deviations from the azimuthal mean flow, $\|u_\theta-\langle u_\theta\rangle\|$, where red (blue) is large (small).}
\end{figure}

\section{Experimental System}\label{sec:es}
The geometry of the TCF apparatus used in this study, illustrated in \autoref{fig:TCF_geometry}, is characterized by two geometric parameters: the radius ratio $\eta = {r_i}/{r_o}$ and the aspect ratio $\Gamma = {h}/({r_o-r_i})$, where $r_i=50$\,mm and $r_o=70.24$\,mm are, respectively, the inner and outer radii of the flow domain. In his seminal paper, Taylor explored the $\Gamma\gg 1$ limit in which boundary effects of the top and bottom end caps are negligible; in contrast, we investigate the small aspect-ratio, $\Gamma = 1$, parameter regime, in which the end effects play a substantial role in driving the fluid motion.

The inner and outer cylindrical walls of the Taylor-Couette apparatus rotate with angular velocities $\omega_i$ and $\omega_o$, respectively, with the top and bottom endcaps co-rotating with the outer wall. This naturally defines two Reynolds numbers
\begin{align*}
    \text{Re}_{i} = \frac{r_{i}\omega_{i}(r_o-r_i)}{\nu} && \text{Re}_{o} = \frac{r_{o}\omega_{o}(r_o-r_i)}{\nu},
\end{align*}
which characterize the external driving of the fluid. Here, $\nu$ is the kinematic viscosity. We investigate the case of counter-rotating boundaries $\text{Re}_o = -200$ and $\text{Re}_i=500$.

The experimental TCF apparatus was constructed to be fully transparent to allow for optical access and, therefore, velocimetry measurements anywhere in the flow domain. Time-resolved azimuthal and radial velocity measurements near the horizontal midplane were sufficient in characterizing the dynamics, despite the flow having nontrivial structure in the axial direction \cite{Crowley2022}. Here, the experimental setup and velocimetry techniques are identical to that of \cite{Crowley2022}; velocity is measured in a horizontal cross section, as depicted in \autoref{fig:TCF_geometry}.

\section{Exact Coherent Structures in TCF}\label{sec:ecstcf}

\begin{table}[]
\centering
\begin{tabular}{c|cc c c}
      & $T$            & $\Phi$  & $N^u$  & $\gamma$  \\ \hline
RPO${}_{1}$ & 0.196 & 1.043 & 9 & $0.0246$\\
RPO${}_{2}$ & 0.177 & 0.856 & 7 & $0.0209$\\
RPO${}_{3}$ & 0.234 & 0.448 & 9 & $0.0260$\\
RPO${}_{4}$ & 0.200 & 0.199 & 8 & $0.0299$\\
RPO${}_{5}$ & 0.422 & 0.443 & 7 & $0.0336$\\
RPO${}_{6}$ & 0.419 & 0.425 & 8 & $0.0358$\\
RPO${}_{7}$ & 0.164 & 0.481 & 8 & $0.0342$\\
RPO${}_{8}$ & 0.215 & 5.799 & 8 & $0.0464$
\end{tabular}
\caption{\label{tab:1} \small Properties of known RPOs in our TCF system at $\text{Re}_i=500$ and $\text{Re}_o=-200$. The table lists the temporal period, $T$; azimuthal shift, $\Phi$; dimension of the unstable manifold of each solution, $N^u$; and escape time, $\gamma$. Both $T$ and $\gamma$ are reported in non-dimensional units, normalized by the viscous timescale $(r_o-r_i)^2/\nu\approx 271$. $N^u$ includes the $2$ marginally stable directions of each solution, along $\tau$ and $\phi$.}
\end{table}

The presence of a continuous (rotational) symmetry in Taylor-Couette flow implies that the most common ECSs will be relative, e.g., traveling waves or relative periodic orbits (RPOs) \cite{Meseguer2009b,Budanur2017,Krygier2021}. 
RPOs are solutions satisfying
\begin{equation}\label{eq:rpo}
    {\bf u}(r,\theta,z,t) = {\bf u}(r,\theta+\Phi,z,t+T),
\end{equation}
for some temporal period $T$ and rotation angle $\Phi$. RPOs are time-periodic solutions in a co-moving reference frame of angular speed $\Phi/T$.  

The solutions listed in \autoref{tab:1} were computed by detecting near-recurrences in  direct numerical simulation of turbulent flow \cite{Crowley2022}. 
Specifically, we identified the minima of the difference $\|{\bf u}(r,\theta,z,t)-{\bf u}(r,\theta+\Phi,z,t+T)\|$ defined in terms of the 2-norm
\begin{align}\label{eq:2norm}
    \|{\bf u}\| = \left[\int_V ({\bf u}\cdot{\bf u})\, dV\right]^\frac{1}{2}.
\end{align}
Here, $V$ corresponds to a thin horizontal slice, $V=[0,2\pi)\times[r_i,r_o]$, centered at $z=0$ ($z=1.5$\, mm) in numerics (experiment). The corresponding tuples $({\bf u},T,\Phi)$ were then provided as initial conditions to a Newton-Krylov solver \cite{KrygierThesis}, which determined eight solutions, $({\bf u}_n, T_n, \Phi_n)_{n=1}^8$, that satisfy both the Navier-Stokes equation and \autoref{eq:rpo} exactly. Further properties, such as their stability, are also summarized in \autoref{tab:1}.

\begin{figure}[]
    \center
    \subfloat[]{\includegraphics[width=0.45\textwidth]{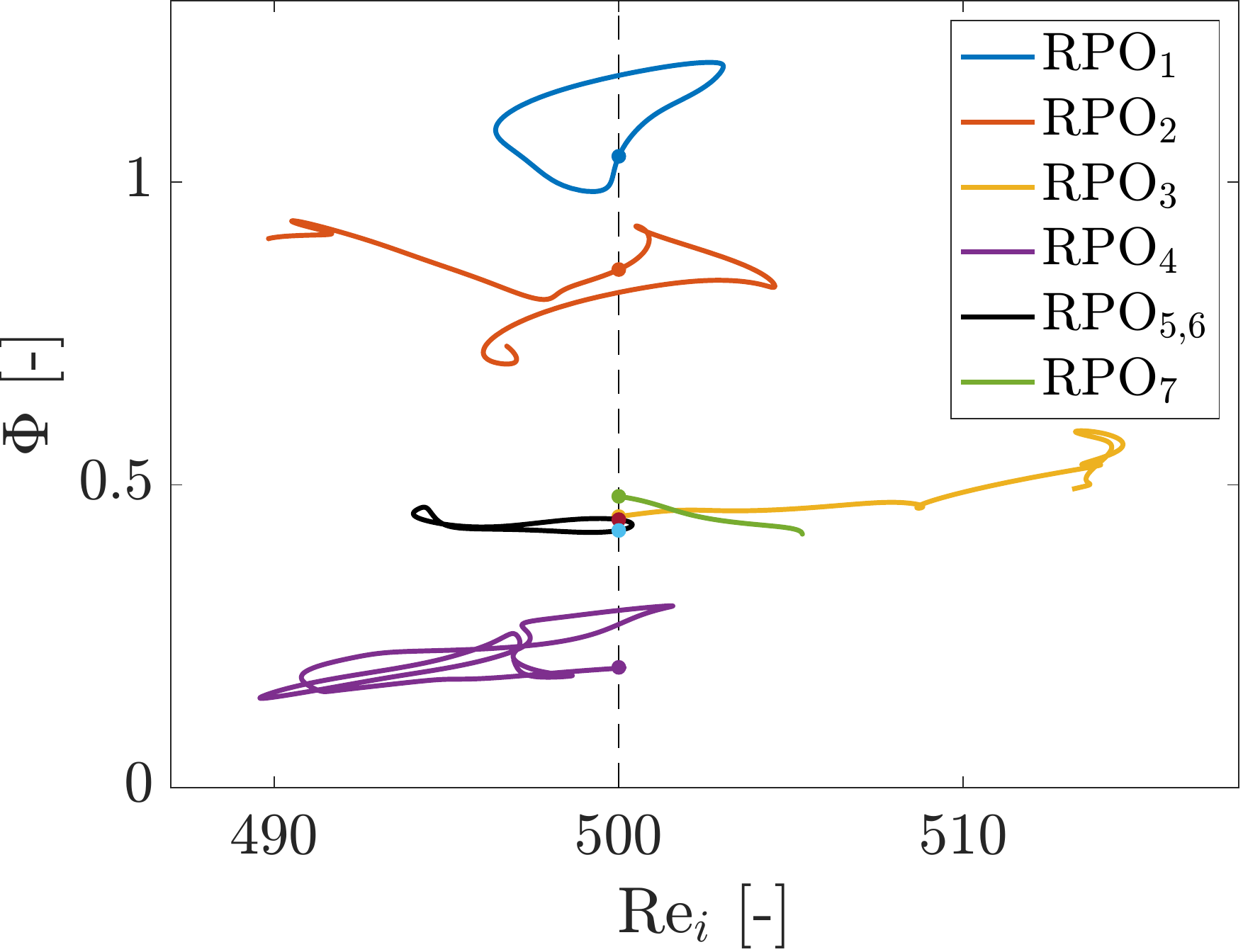}} \hspace{0.5cm}
    \subfloat[]{\includegraphics[width=0.47\textwidth]{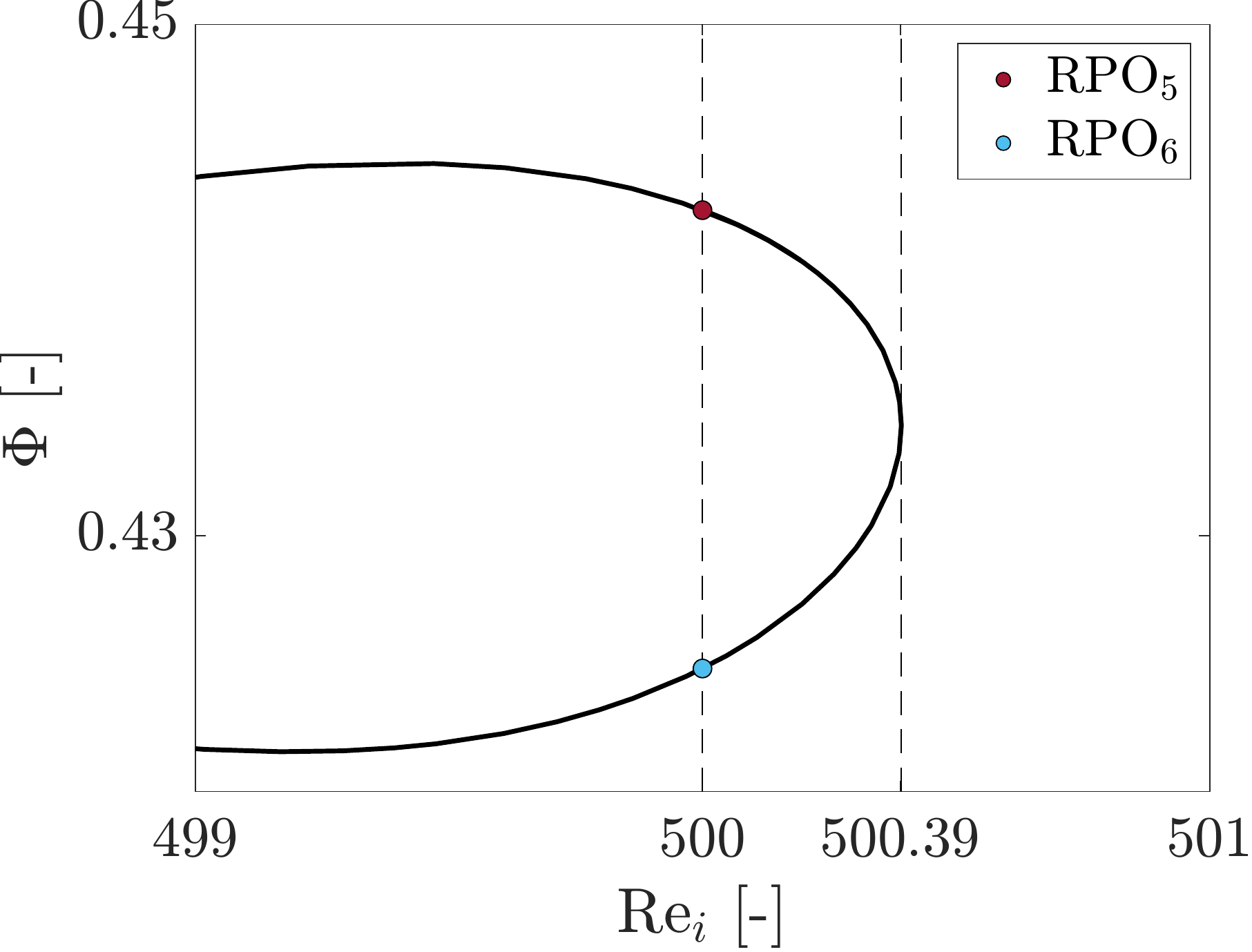}}
    \caption{\label{fig:continuation} (a) Solution branches associated with RPOs listed in \autoref{tab:1} obtained by continuation in Re$_i$. For each, rotation angle $\Phi$ is plotted as a function of Re$_i$, with the initial solution at Re$_i=500$ plotted as a solid marker. RPO$_5$ and RPO$_6$ lie on the same continuation curve, as shown in panel (b), and are born from a bifurcation at Re$_i = 500.39$.
    }
\end{figure}

Additional ECSs can be found by numerical continuation of the eight original RPOs in any of the parameters (Re$_i$, Re$_o$, $\Gamma$, $\eta$) \cite{KrygierThesis}. For instance, continuation in Re$_i$ yields solution branches that commonly turn around at saddle-node bifurcations, returning to the original value of Re$_i$. This procedure either connects the original RPOs (e.g., RPO$_5$ and RPO$_6$ as shown in Figure \ref{fig:continuation}) or defines new ones. Note that, aside from such rare exceptions, the properties of the RPOs are not particularly sensitive to the choice of parameters. This enables comparison to experiment that has an uncertainty with how parameters (mainly Re$_i$ and Re$_o$) are set or measured.


In the presence of continuous symmetry, existence of an ECS implies the existence of a continuum of symmetry-related copies of that ECS. 
In particular, for an RPO described by the velocity field ${\bf u}_n(r,\theta,z,t)$, the symmetry-related copies ${\bf u}_n(r,\theta+\phi,z,t+\tau)$ also satisfy both the governing equations and the condition \eqref{eq:rpo} for any $\tau$
and $\phi$,
which are most naturally thought of as the group parameters describing translational symmetry in time and angular orientation, respectively.
Of course, not all symmetry-related solutions are necessarily distinct, e.g., $\tau=kT_n$ and $\phi=k\Phi_n$ with any integer $k$ yield identical flows related by a shift in the temporal origin.

Consider the flow with an initial condition ${\bf u}_n(r,\theta,z,0)$. In the typical case when $\Phi_n$ is an irrational multiple of $2\pi$, this flow is quasi-periodic in the laboratory frame and is characterized by two frequencies, $\Omega_1=2\pi/T$ and $\Omega_2=\Phi_n/T_n$. The condition \eqref{eq:rpo} ensures that this flow can be represented by properly rotated copies of the RPO ${\bf u}_n(r,\theta,z,\tau)$ with $0\le \tau<T_n$, so we only need to compute one period to represent the flow at all times. 
Topologically, a quasi-periodic flow corresponds to a torus which, for the case of an RPO, can be fully parameterized by the velocity field ${\bf u}_n(r,\theta+\phi,z,t+\tau)$ with $\tau\in[0,T_n)$ 
and $\phi\in [0,2\pi)$. 
As a result, for any $\Phi_n$ which is an irrational multiple of $2\pi$, the set of all distinct translations span the surface of a torus, and all points on this torus may be parametrized by the cyclic temporal coordinate $\tau$ and angular coordinate $\phi$. On the surface of this torus, temporal evolution is fundamentally simple in the $(\tau,\phi)$ parameterization: 
\begin{align}\label{eq:phases}
    \tau &= t+\tau_0-kT_n \nonumber\\
    \phi &= \phi_0+k\Phi_n,
\end{align}
for $t\in[kT_n,(k+1)T_n)$. This parameterization of the torus is analogous to the one used by \cite{Suri2018} to parameterize unstable manifolds. 

\section{State Space Geometry and Shadowing}\label{sec:ssg}
For an observer, an evolving fluid flow is most naturally described by a time-dependent velocity field. Geometrically, the same flow can be described by a one-dimensional trajectory in a high-dimensional state space. State space can be constructed from the full velocity field with any smooth, invertible mapping, such as a Fourier or POD decomposition. Regardless of the choice of mapping, at every instant in time, the entire flow field corresponds to a single point in the state space. As time evolves, this point traces out a curve which, for a turbulent flow, has a very complicated and highly tangled shape.

These tangled trajectories are direct analogues of chaotic solutions to systems of nonlinear differential equations such as the well-known Lorenz \cite{lorenz1963} or R{\"o}ssler \cite{Rossler1976} systems. As a result of dissipation,  chaotic trajectories are attracted, and eventually confined to, 
a lower-dimensional manifold known as the chaotic attractor. In addition to  chaotic trajectories, the attractor also contains an infinite number of unstable recurrent solutions, such as periodic orbits. Quite analogously, the trajectory describing a turbulent flow is also confined to a relatively low-dimensional manifold which contains unstable recurrent solutions (ECSs) of the Navier-Stokes equations. 
Just as in the low-dimensional setting, ECSs represent low-dimensional invariant sets: e.g., equilibria correspond to points (0-dimensional sets), traveling waves and time-periodic solutions correspond to loops (1-dimensional sets), and quasi-periodic solutions such as RPOs correspond to tori (2-dimensional sets).

The deterministic nature of the governing equations implies that flows with different yet sufficiently close initial conditions will remain close and will evolve similarly over a period of time. In particular, if turbulence comes close enough to one of the recurrent solutions, the two flows will co-evolve or shadow each other 
until they eventually diverge. However, the manifold containing the turbulent trajectory is relatively high-dimensional; the largest $N^u$ provides a lower bound to the dimensionality of this manifold, see \autoref{tab:1}.
Hence, it is {\it a priori} unclear whether in experiment turbulence comes sufficiently close to, and therefore shadows, any one of the known recurrent solutions on accessible time scales \cite{Crowley2022, Suri2020}.    

A geometrical representation of turbulence
is particularly helpful for both understanding and quantifying the 
relation 
between turbulence
and recurrent solutions. Dynamical similarity between any two evolving flows is described by the 
shape and proximity of the corresponding trajectories in state space. 
Points that are close in state space, as characterized by Euclidean distance $\|{\bf u}-{\bf v}\|$, correspond to flows ${\bf u}$ and ${\bf v}$ with similar velocity fields.
Similarity in the shape of (portions of) a pair of trajectories in the state space implies qualitative similarity in the evolution of the corresponding flows.
Quantitative similarity additionally requires that the state space speed be similar for the two trajectories \cite{Suri2020}. 


\begin{figure}[]
    \center
    \subfloat[]{\includegraphics[width=.4\textwidth]{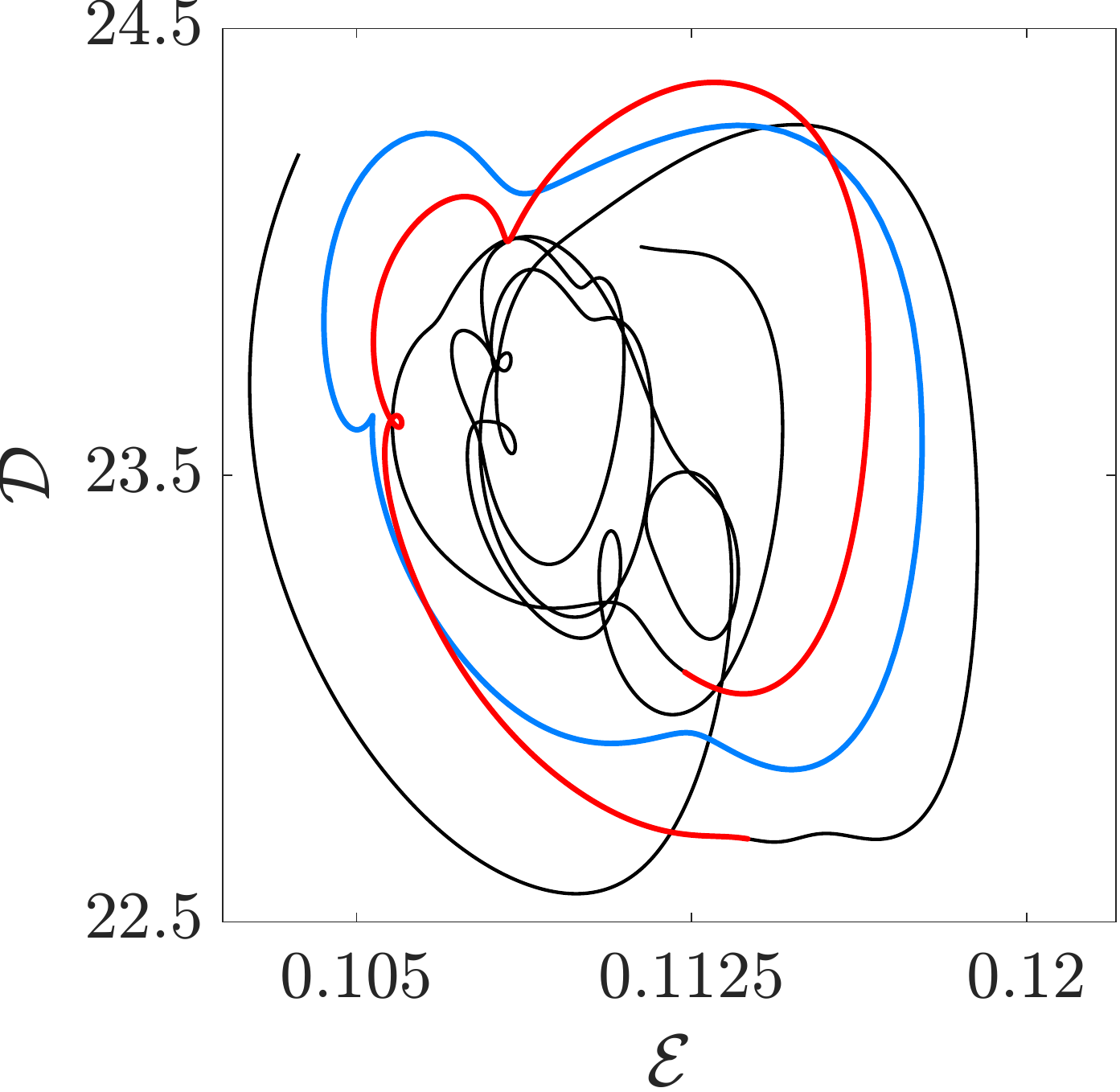}} \hspace{0.5cm}
    \subfloat[]{\includegraphics[width=.418\textwidth]{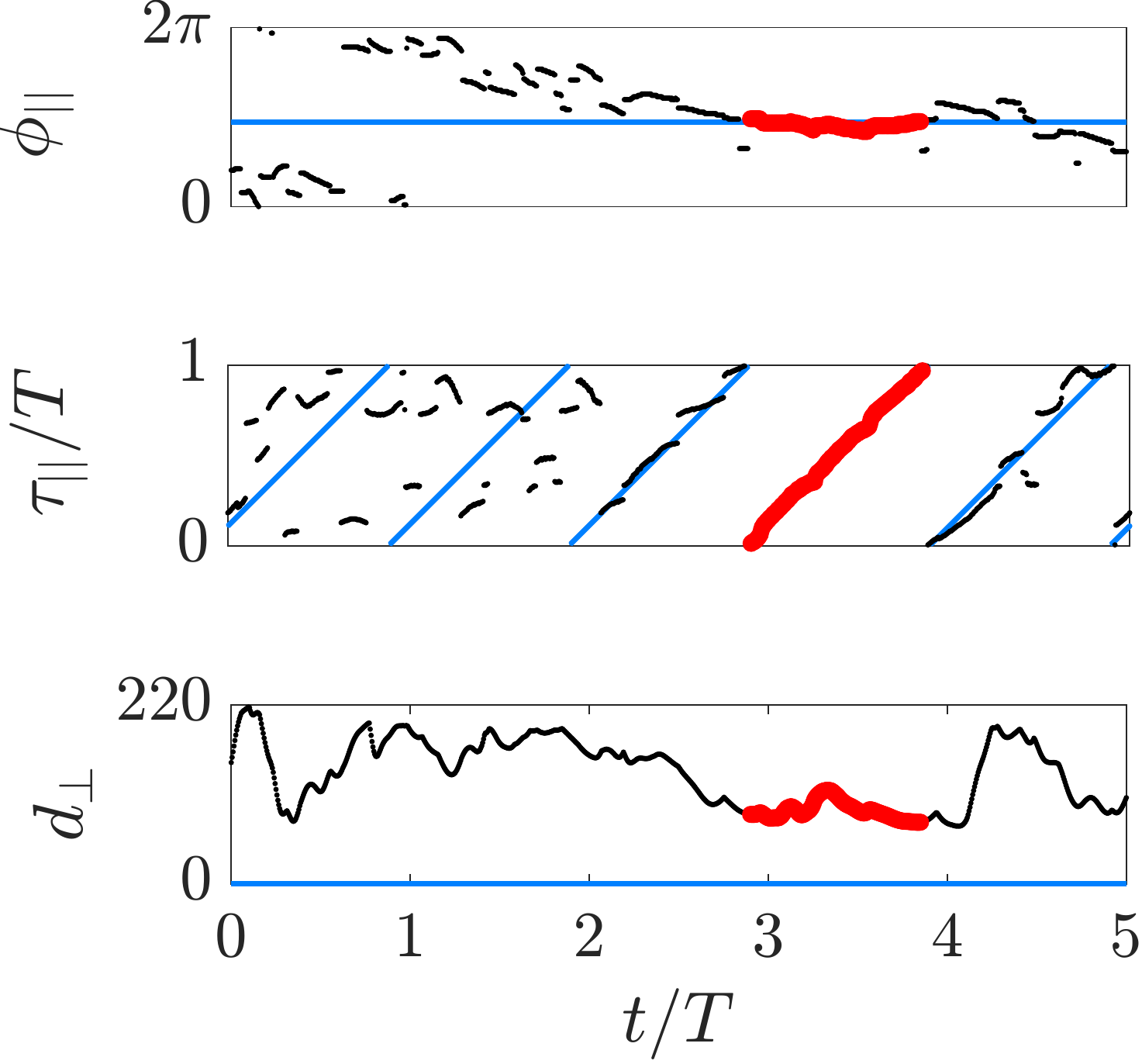}}
    \caption{\label{fig:shadowing_tutorial} An example of shadowing using different projections of state space.  (a) A traditional projection onto the two-dimensional subspace spanned by the energy density $\mathcal{E}$ and dissipation rate $\mathcal{D}$. (b) The same interval of time, projected onto the local coordinates: azimuthal phase $\phi_{||}$, temporal phase $\tau_{||}$, and distance $d_\perp$. In both panels, RPO$_1$ is shown in blue and turbulent flow is shown in black (red) outside (inside) the shadowing interval. The interval shown corresponds to the shadowing event at $t\approx 17$ min in \autoref{fig:BarcodePlots}.} 
\end{figure}

While comparison of a pair of flow snapshots is straightforward, quantifying the similarity of evolving flows is far less trivial. Generic low-dimensional projections frequently used in the literature can be misleading. Consider, for instance, the projection onto the subspace spanned by two common global observables, the energy density
\begin{equation}
    \mathcal{E} = \frac{1}{\text{Re}^2V}\int_{V} {\bf u}^2 \, dV
\end{equation}
and the rate of energy dissipation
\begin{align}
    \mathcal{D} = \frac{1}{\text{Re}^2V}\int_{V} \boldsymbol\omega^2 \, dV, 
\end{align}
normalized here by the characteristic velocity scale $Re=|Re_i-Re_o|/2$. Shown in Figure \ref{fig:shadowing_tutorial}(a), this projection suggests that a portion of turbulent trajectory (shown in red) has a shape qualitatively similar to that describing RPO$_1$ (shown in blue). The corresponding turbulent flow might indeed be similar to this RPO with a very particular orientation. However, the exact same projection would also describe an RPO with a completely different orientation that is not at all similar to the turbulent flow (i.e., lies far from it in the full state space). 


A more appropriate choice of coordinates is local rather than global and is informed by the symmetries of the problem. In the neighborhood of an RPO, every trajectory ${\bf u}(t)$ can be well-parameterized \cite{Krygier2021} using the distance transverse to the torus
$$d_\perp(t)=\min_{\tau,\phi}d_n(t,\tau,\phi)$$ 
and the two phases along its surface,
$$\{\tau_{\|}(t),\phi_{\|}(t)\}={\rm arg\,min}_{\tau,\phi}d_n(t,\tau,\phi)$$
defined relative to the RPO, where \begin{equation}\label{eq:dn}
d_n(t,\tau,\phi)=\|{\bf u}(r,\theta,z,t)-{\bf u}_n(r,\theta+\phi,z,\tau)\|.
\end{equation}
 We will drop the subscript $n$ when it is clear which solution is being considered. 

For a trajectory lying on the torus representing the RPO, $d_\perp(t)=0$ and its evolution is described by \autoref{eq:phases} exactly. For $d_\perp(t)$-small, one would expect \autoref{eq:phases} to be satisfied approximately. Indeed, this is what we find for the segment of the turbulent flow shown in red in \autoref{fig:shadowing_tutorial}. As panel (b) illustrates, over an interval comparable to one period of RPO$_1$, the temporal phase $\tau_{\|}(t)$ indeed faithfully follows a diagonal, straight line, indicating that turbulent flow evolves at the same rate as the RPO. The azimuthal phase $\phi_{\|}(t)$ remains nearly constant over the same interval.

\begin{figure}[]
    \center
    \includegraphics[width=\textwidth]{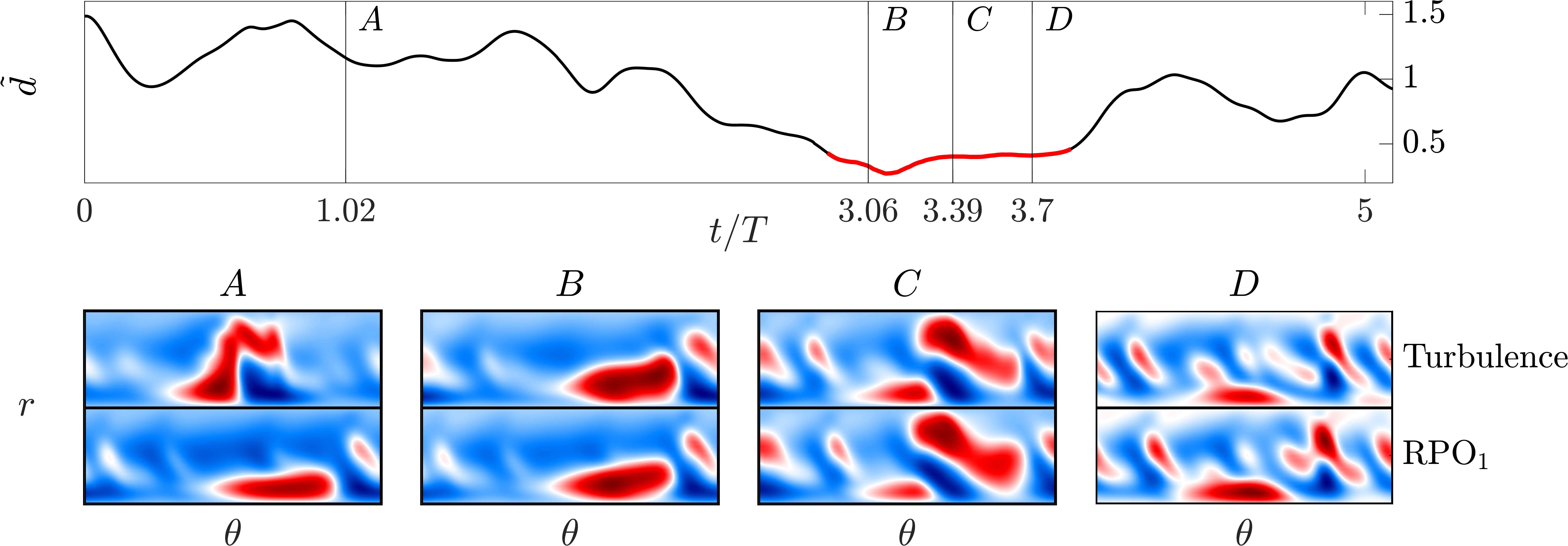}
    \caption{\label{fig:ShadwingQuality} Comparison of the flow fields representing turbulent flow and relative periodic orbit (RPO$_1$) during the same interval illustrated in \autoref{fig:shadowing_tutorial}. (Top) The distance $\tilde{d}$ to the RPO with the shadowing portion is shown in red. (Bottom) The flow fields corresponding to the time instances are labeled with vertical lines in the top panel. Here and below, the mean-subtracted azimuthal velocity, $u_\theta-\langle u_\theta\rangle$, is plotted for turbulent flow and the corresponding RPO, where red (blue) is positive (negative).
    Each rectangular box represents the annular region spanned by $r$ and $\theta$, where $r_i$ is on the bottom and $r_o$ is on the top. 
    In \autoref{fig:BarcodePlots}(a), this event corresponds to the minimum at $t\approx36$ minutes.}
\end{figure}

\section{Detecting Shadowing in Experiment}\label{sec:detect}

Computation of the phases corresponding to the minima of the distance $d(t,\tau,\phi)$ becomes unreliable in experiment due to unavoidable noise. To address this, we instead quantify dynamical similarity by time-averaging the distance over a finite time interval $t\in[t_0-I/2,t_0+I/2]$  
\begin{equation}\label{eq:UnNormedDistTime}
    \bar{d}(t_0,\tau_0,\phi_0) = \frac{1}{I}\int_{t_0-I/2}^{t_0+I/2}
    d(t,\tau(t),\phi(t))\ dt,
\end{equation}
where the two phases are assumed to satisfy \autoref{eq:phases} at all times.
Shadowing of the respective RPO over this interval is expected for $\bar{d}$ sufficiently small and, indeed, this is what we find, as illustrated in \autoref{fig:ShadwingQuality}.

The natural time scale associated with a close pass to an RPO is given by the inverse of the escape rate,
$$\gamma_n = \left[\sum_i \text{Re}(\lambda_{n,i})\right]^{-1},$$
where $\lambda_{n,i}$ are the unstable Floquet exponents of RPO$_n$. Hence we take $I=\gamma_n$ for detecting shadowing of the corresponding RPO. As \autoref{tab:1} shows, these escape times tend to be quite short compared to the periods of the respective RPOs. The shadowing event discussed in the previous section is particularly long: its duration is comparable to one entire period of the RPO (almost $10I$). 



\begin{figure}[h]
    \center
    \subfloat[]{\includegraphics[width=0.5\textwidth]{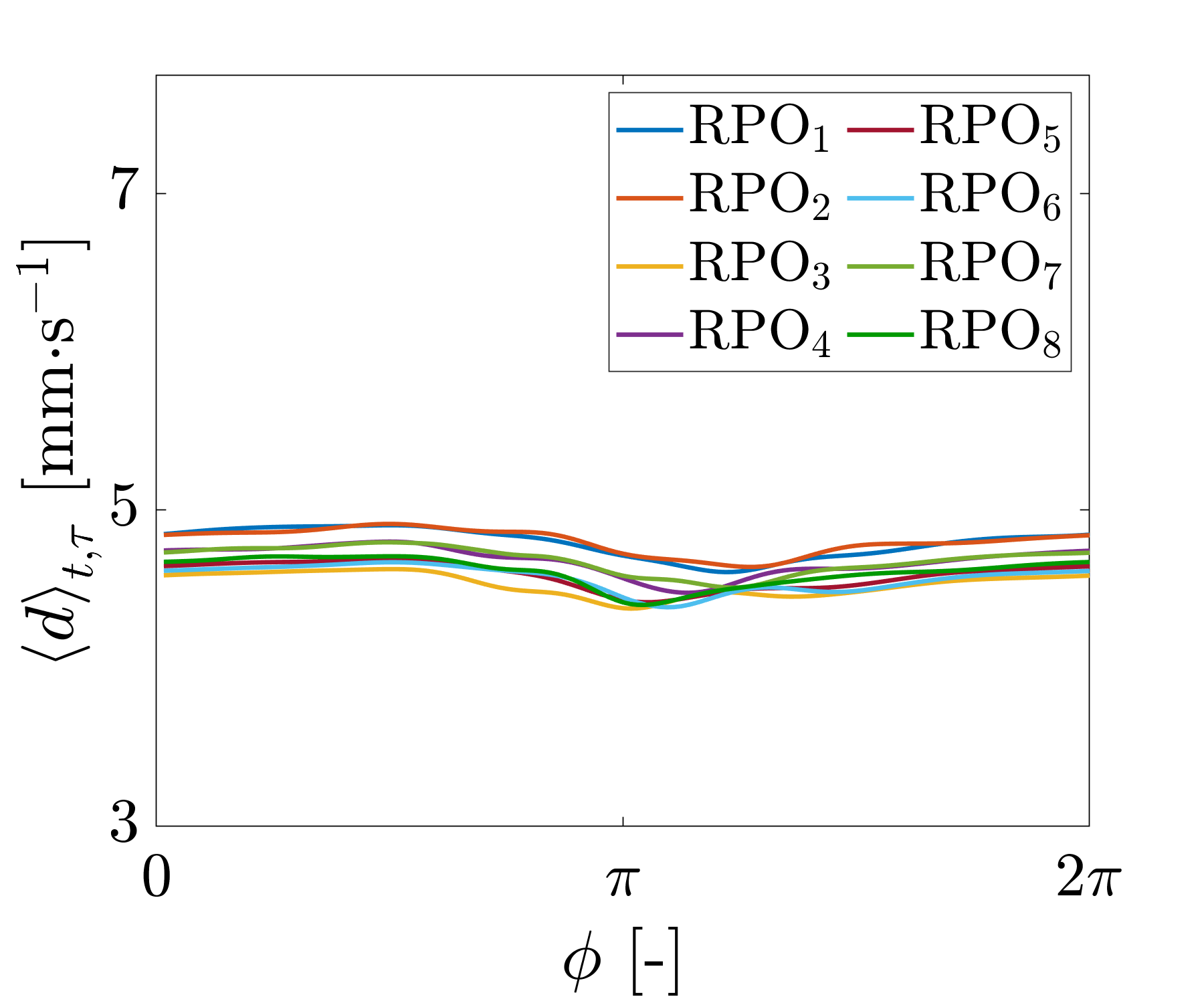}}
    \subfloat[]{\includegraphics[width=0.5\textwidth]{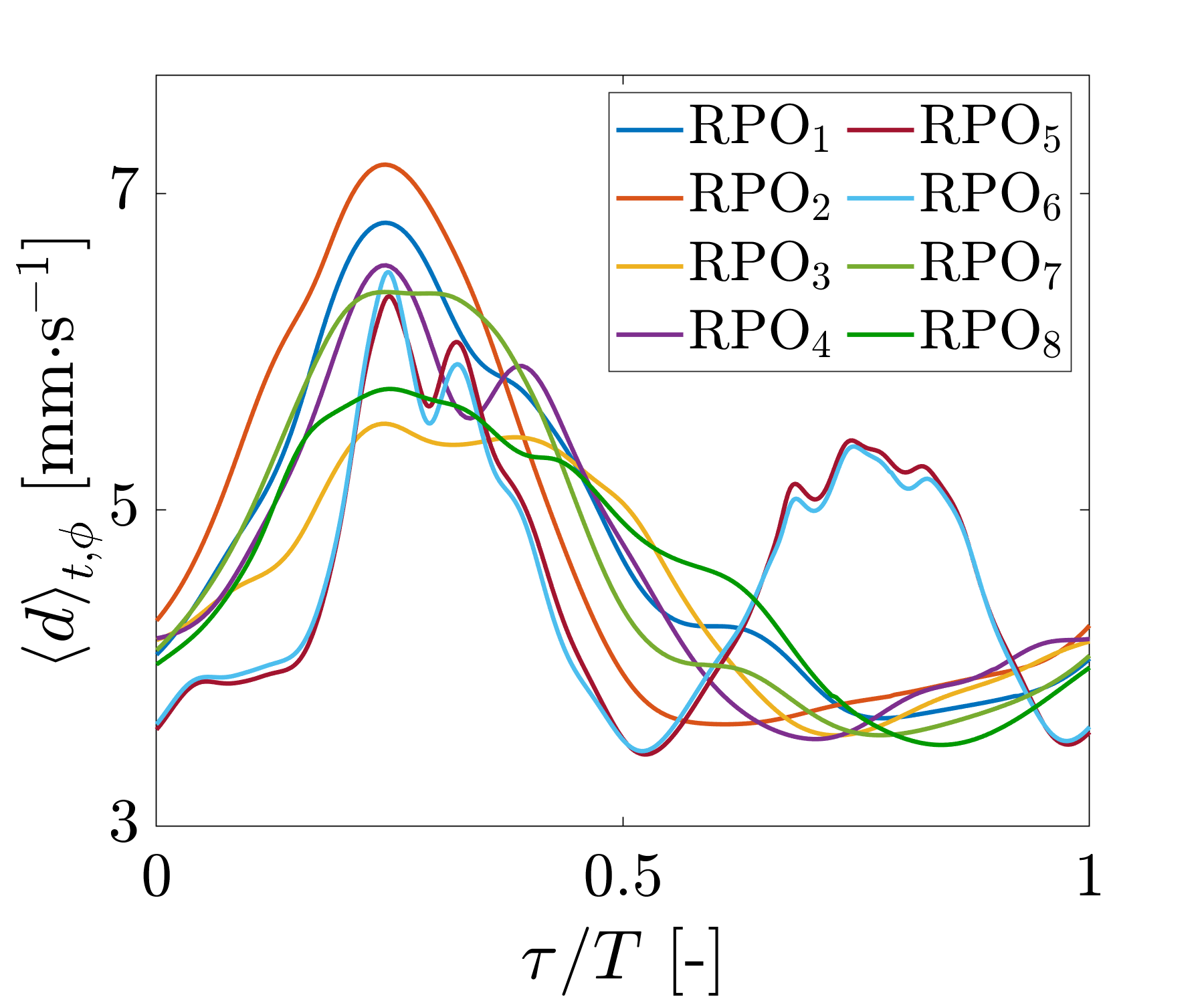}}
    \caption{\label{fig:NormalizeTheCR} Phase dependence of the distance defined in \autoref{eq:dn} averaged (a) over $t$ and $\tau$ and (b) over $t$ and $\phi$. Here, $\langle \ \cdot \ \rangle_x$ denotes an average over $x$. 
    }
\end{figure}

To characterize $\bar{d}$ as "small" or "large," a distance scale is needed. Generically, a characteristic distance scale, $D$, can vary substantially throughout state space, as the density of trajectories in state space itself varies. This characteristic distance can even vary along a single recurrent solution. As a result, intervals of shadowing that occur in regions where $D$ is large will result in a larger value of $\bar{d}$ compared with shadowing events of similar quality in regions where $D$ is small. Thus, to compensate for this intrinsic variation in distance within state space, we define a normalized average distance 
\begin{equation}\label{eq:NormedDistConvolve}
    \tilde{d}(t_0,\tau_0,\phi_0) = 
    \frac{1}{I}\int_0^{T_n}d\tau \int_0^{2\pi}d\phi
    \int_{t_0-I/2}^{t_0+I/2}  \frac{d(t,\tau(t),\phi(t))}{D(\tau(t),\phi(t))}\,dt 
\end{equation}
where $D(\tau,\theta)$ is the characteristic distance scale of state space in the vicinity of the RPO, ${\bf u}(\tau,\theta)$. In general, $D$ may depend on $\theta$. However, in TCF, rotational symmetry ensures that $D$ is independent of $\theta$ (see \autoref{fig:NormalizeTheCR}(a)). On the other hand, the dependence on $\tau$ is quite pronounced for all RPOs, as \autoref{fig:NormalizeTheCR}(b) illustrates. Thus, we define the characteristic distance scale 
\begin{equation}\label{eq:normalization}
    D(\tau) = \left[\frac{1}{2\pi T}
    \int_0^{2\pi}d\theta\int_0^T \frac{1}{d(t,\tau,\theta)}\ dt\right]^{-1},
\end{equation}
as the harmonic mean distance between turbulent flow and a given RPO, where $T$ is the duration of the turbulent trajectory. A harmonic mean weighs small distances more than would an arithmetic mean, which provides a more local measure of the characteristic distance to the RPO. 
 

The normalized distance $\tilde d$ provides a more uniform measure of the quality of shadowing. Low values correspond to intervals when both the turbulent flow field and its time evolution are well captured by the specific recurrent solution. For larger values of $\tilde d$, the apparent similarity of the two flow fields and dynamics is less striking. The degree of similarity between the two flows that defines shadowing is however subjective. As \autoref{fig:ShadwingQuality} illustrates, $\tilde d = 0.5$ gives a threshold of 
visual 
similarity between the flow fields in numerical simulation. Experimental noise increases $d$ and thereby raises the experimental threshold to $\tilde d=0.65$. 


\section{Results}\label{sec:results}

\begin{figure}[]
    \center
    \subfloat[]{\includegraphics[width=\textwidth]{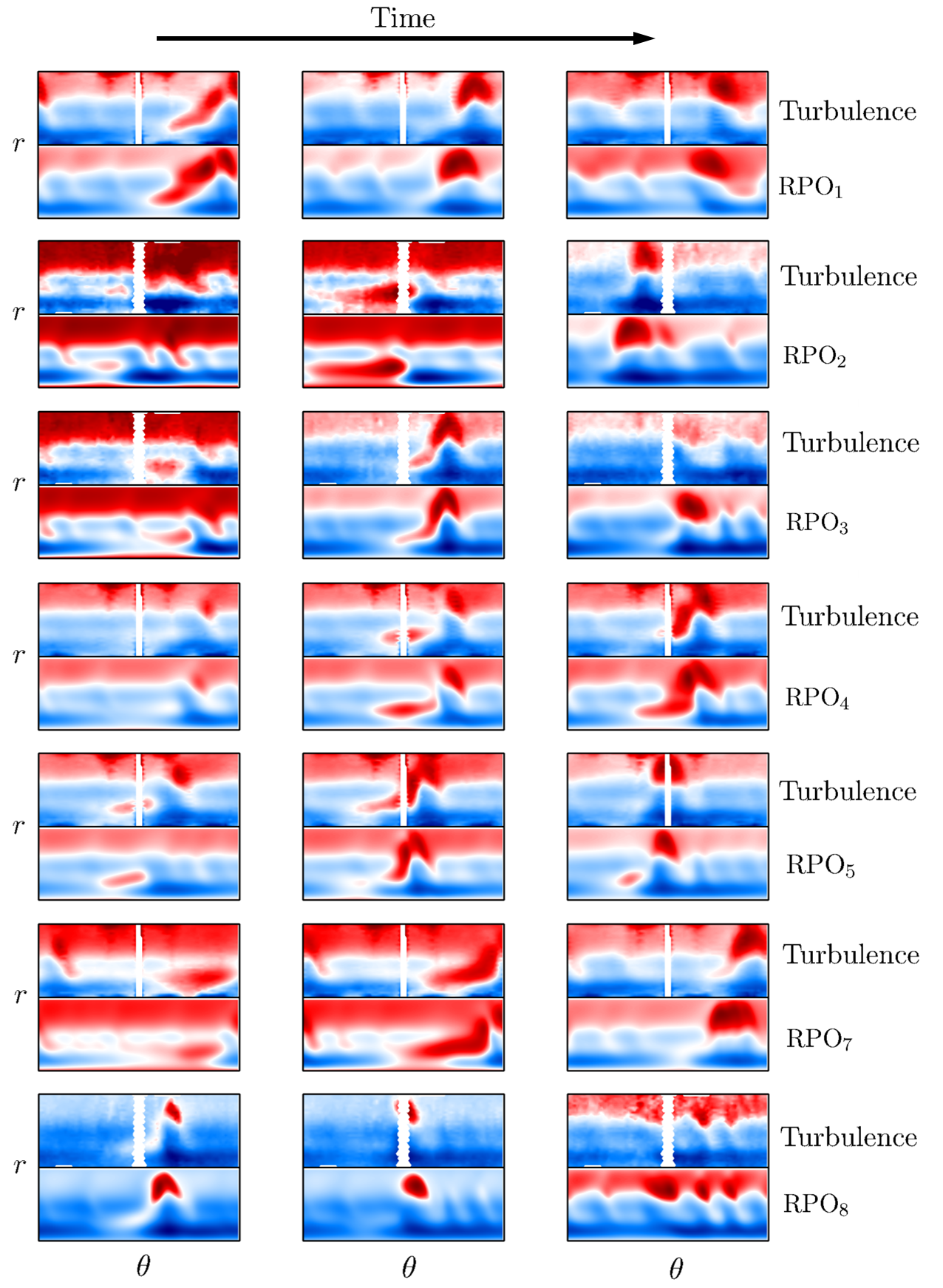}}
    \caption{\label{fig:RealSpaceShadowing} Experimental observations of turbulence shadowing each of the known recurrent solutions (due to the similarity of RPO$_5$ and RPO$_6$, only RPO$_5$ is shown). The vertical band represents experimental data missing because of nonuniform illumination. In each example, time advances from left to right, where the separation between each column is approximately 6.4 seconds, or about one eighth of the typical period. No two events shown here are simultaneous.  }
\end{figure}
\begin{figure}[]
    \center
    \subfloat[]{\includegraphics[width=1\textwidth]{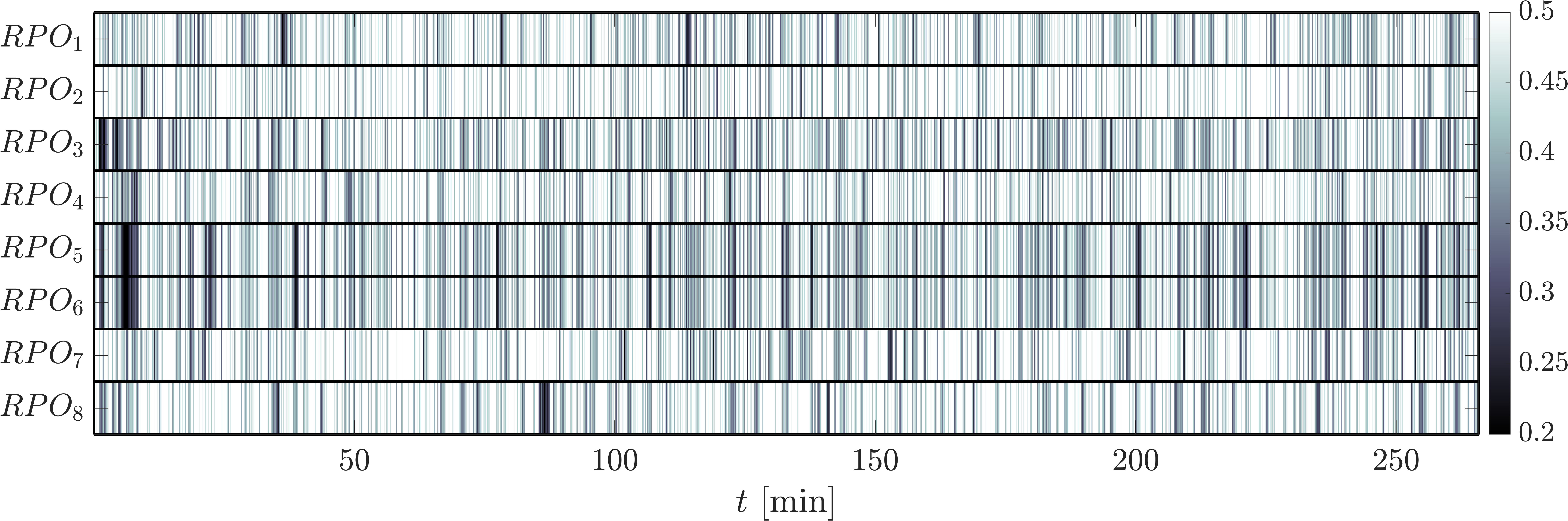}}\\ 
    \subfloat[]{\includegraphics[width=0.38\textwidth]{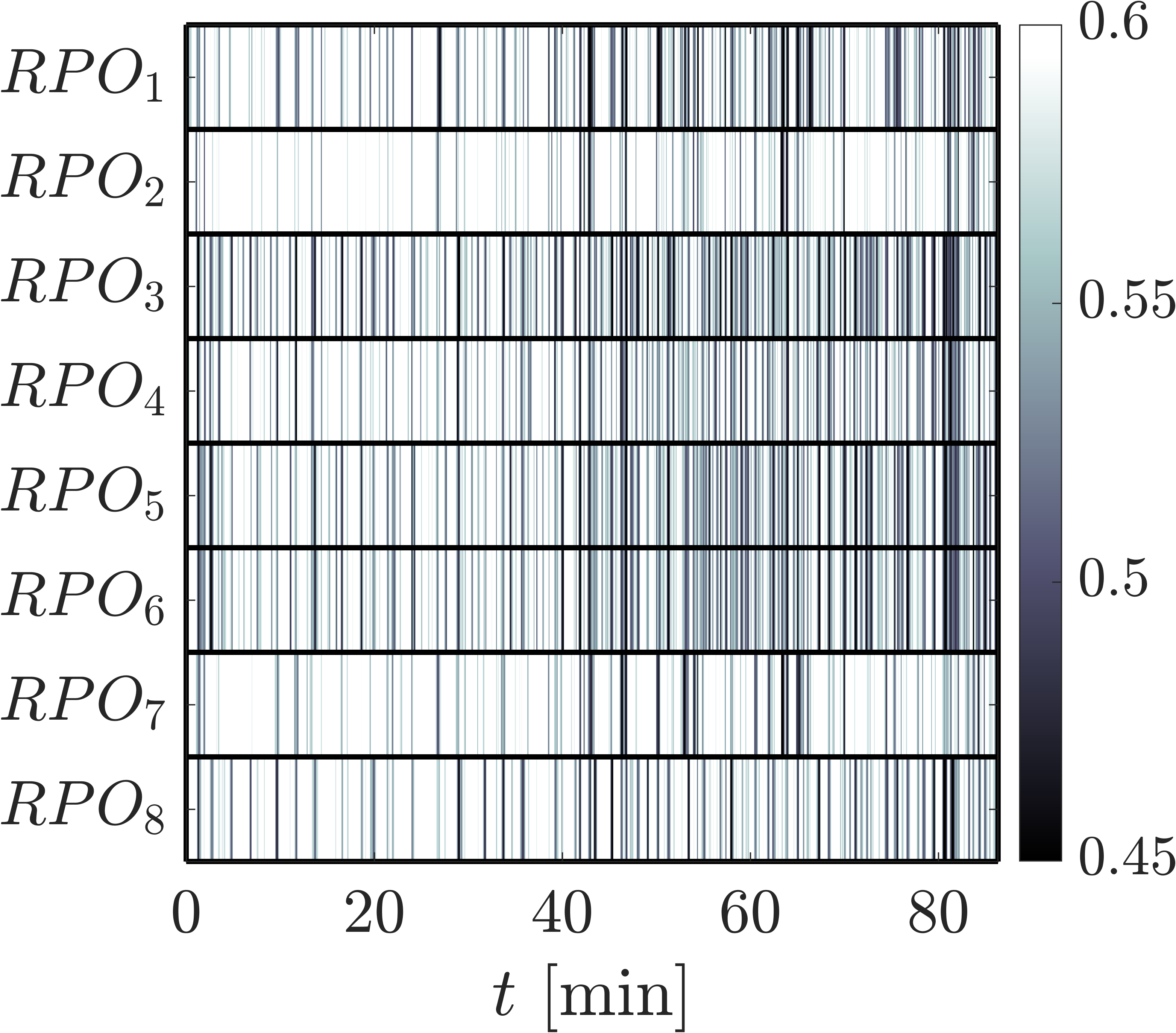}}
    \caption{\label{fig:BarcodePlots} Turbulence frequently shadows relative periodic orbits (RPOs). For in RPO, $\tilde{d}$ is plotted as a function of time using a color gradient. Black vertical lines indicate shadowing events -- time intervals during which RPOs are being tracked by turbulence obtained via (a) numerical simulation and (b) experiment. Darker intervals correspond to higher quality shadowing events. The experimental time-series depicts the concatenation of two concomitant data sets, joined at $t\approx 40$ min. A slight difference in noise floor is notable between data sets. } 
\end{figure}

In a dynamical systems framework pioneered by Hopf, turbulence is described by a sequence of visits to the neighborhoods of different recurrent solutions. In support of this picture, we find that, in both numerics and experiment, turbulence shadows our library of recurrent solutions sequentially and that every known solution is shadowed. Representative shadowing events in experiment are presented in \autoref{fig:RealSpaceShadowing} for the distinct RPOs $1$-$5$,$7$, and $8$. 

Despite the relatively small library of RPOs, shadowing is observed quite often, as demonstrated by \autoref{fig:BarcodePlots}. In simulation, turbulence is found to be shadowing at least one solution, $\min_n\tilde{ d}_n(t)<0.5$, about 33\% of the time. This surprising fact suggests that a semi-quantitative dynamical description of turbulence may require only $\mathcal{O}(10)$ recurrent solutions. The accuracy of this dynamical description is directly related to the choice of the distance threshold. Our RPO library provides a relatively coarse partitioning of the state space into the respective neighborhoods. A finer partition corresponding to a lower value of $\tilde{d}$ will require computing a larger library of recurrent solutions, not necessarily all RPOs.

\begin{figure}[]
    \center
    \subfloat[]{\includegraphics[width=1\textwidth]{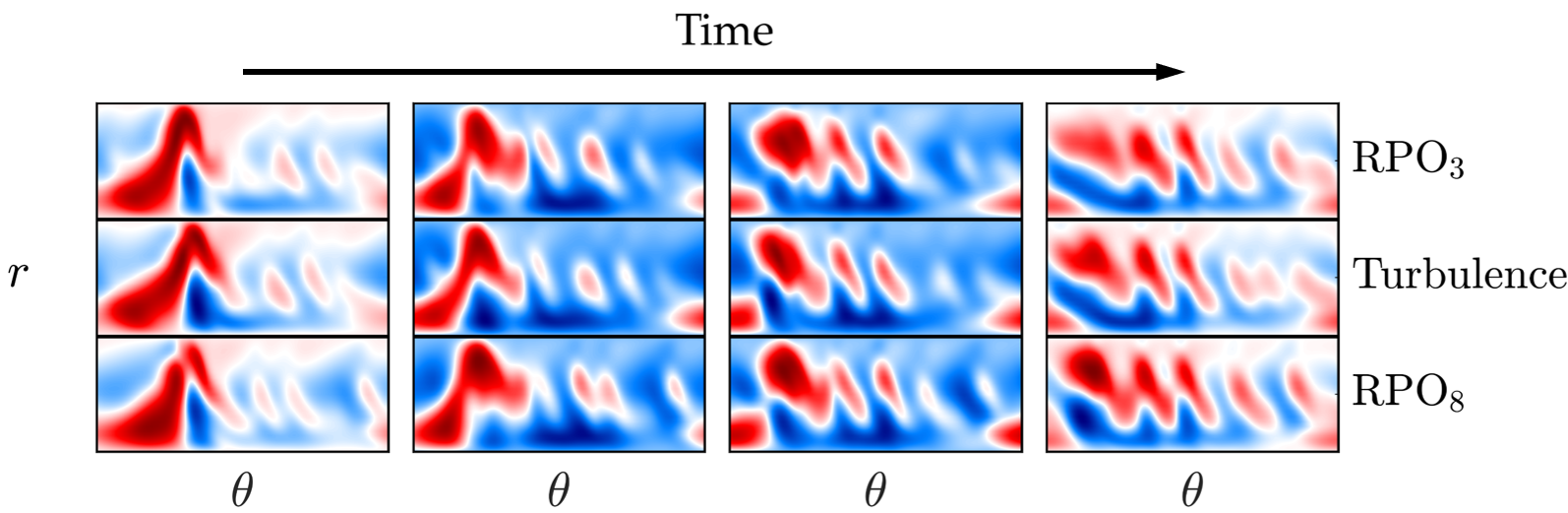}}
    \caption{\label{fig:MultiShadowing} Turbulence (middle) shadows both RPOs 3 (top) and 8 (bottom) for $t\in (15, 40)\ [\text{sec}]$, with panels evolving in time from left to right. The inner three rows display shadowing between all three solutions. Notice that, when turbulence shadows multiple solutions, it resembles all solutions it shadows. During such intervals, the recurrent solutions may also be considered to shadow each other. }
\end{figure}

Another surprising result, given the small size of the RPO library, is that turbulence often shadows more than one solution at once. In contrast to Hopf's conjecture which implies that turbulence is described by one particular recurrent solution at any given instant, we find instead that turbulence may actually be described by multiple solutions, as suggested by \autoref{fig:BarcodePlots} and evidenced by \autoref{fig:MultiShadowing}.

\begin{figure}[]
    \center
    \subfloat[]{\includegraphics[width=0.45\textwidth]{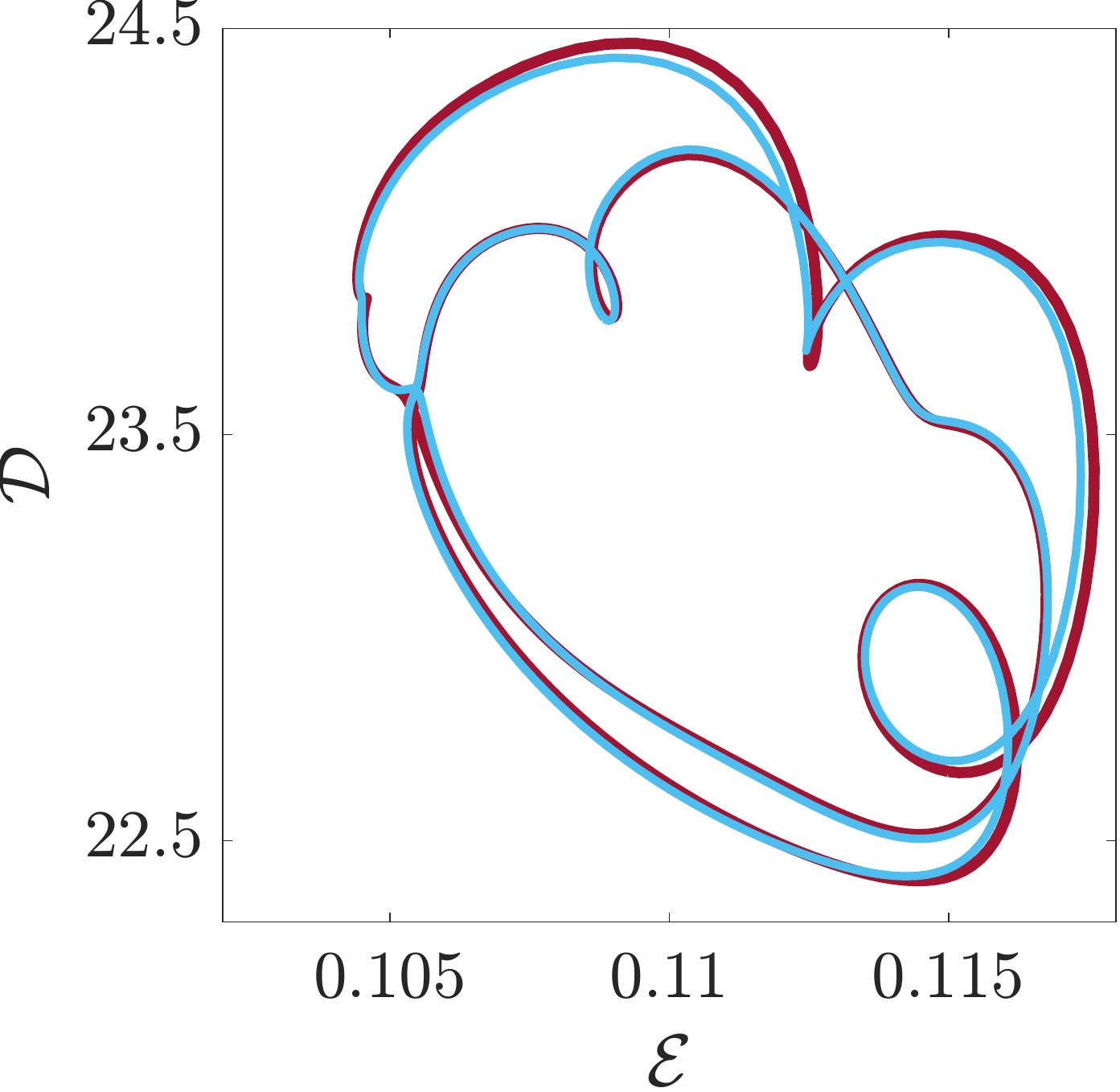}} \hspace{0.4cm}
    \subfloat[]{\includegraphics[width=0.45\textwidth]{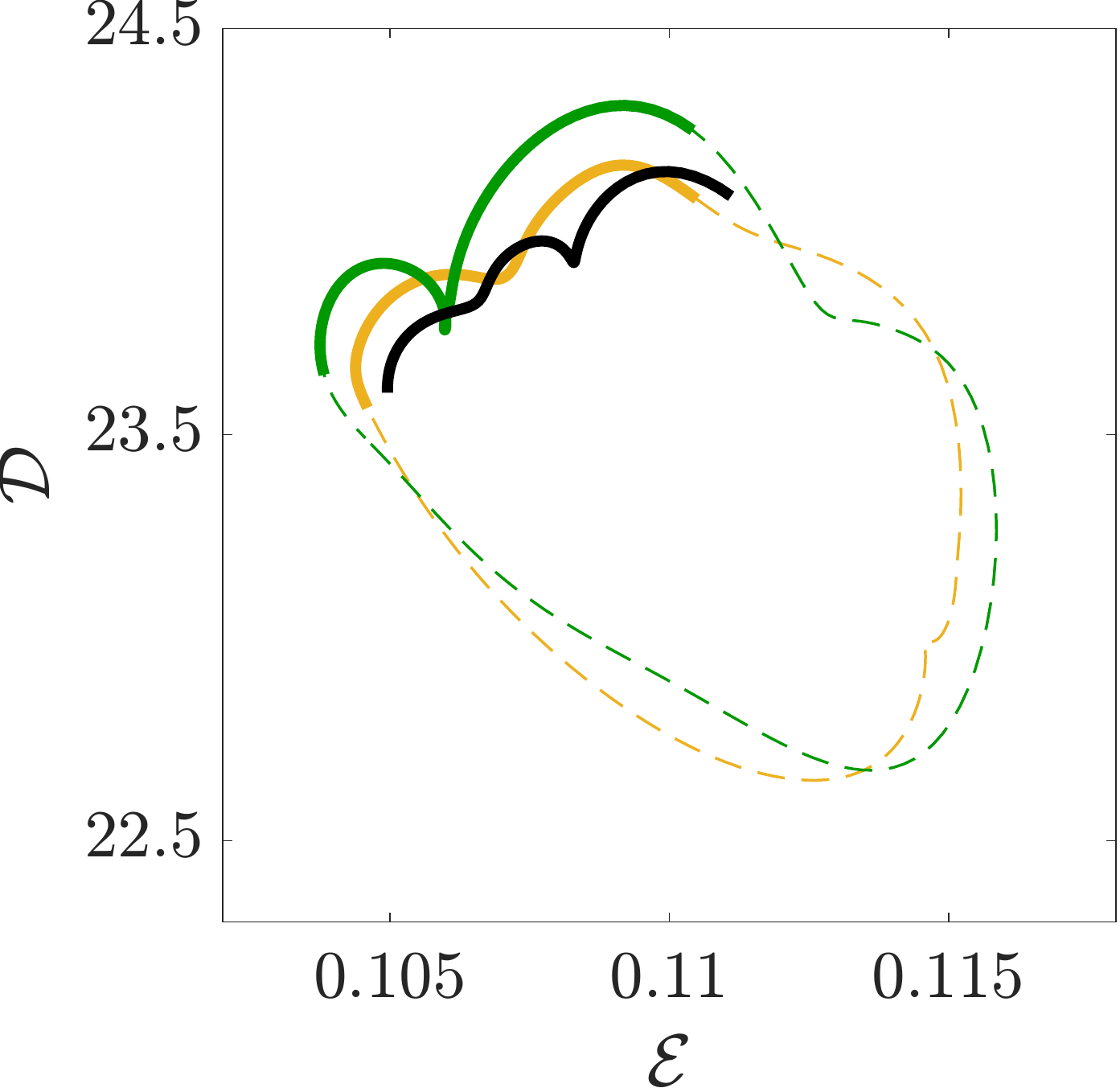}}
    \caption{\label{fig:statespacemultishadow} Low dimensional projections of (a) RPO$_5$ (deep red) and RPO$_6$ (light blue) and (b) RPO$_3$ (yellow) and RPO$_8$ (green) in the statespace. Volumetric kinetic energy and dissipation of the velocity field are used as coordinates. In (b) the intervals of each RPO that are displayed in \autoref{fig:MultiShadowing} are bolded and the portion of turbulent trajectory shadowing these segments is overlaid in black.  }
\end{figure}

Most trivially, one expects turbulence to shadow multiple solutions if these solutions are born from a nearby bifurcation in parameter space and, consequently, are themselves almost indistinguishable. This is the case for RPO$_5$ and RPO$_6$ in our study which are related via a saddle-node bifurcation (see \autoref{fig:continuation}). These two solutions are almost always shadowed simultaneously, which is unsurprising given their proximity in state space, as can be seen in \autoref{fig:statespacemultishadow}(a). 

However, it is also possible for turbulence to simultaneously shadow RPOs which are not related through a nearby bifurcation, as long as portions of these RPOs lie close in state space. \autoref{fig:MultiShadowing} illustrates an example of this simultaneous shadowing in physical space; both RPO$_3$ and RPO$_8$ describe the spatial structure of turbulent flow and its evolution over an interval of $30$ seconds. In this interval, turbulence is inside the neighborhoods of both RPO$_3$ and RPO$_8$, as illustrated by \autoref{fig:statespacemultishadow}(b). In fact, RPO$_3$ and RPO$_8$ effectively shadow each other during this interval as well. 


\section{Conclusion}\label{sec:conc}

    
This study represents one of the first steps in fleshing out Hopf's conjecture that turbulence can be viewed as a deterministic walk between neighborhoods of distinct unstable solutions of the governing equations. 
In small-aspect-ratio counter-rotating Taylor-Couette flow, a number of such solutions have been found previously; they all correspond to RPOs \cite{Crowley2022}. 
We firmly establish that turbulence shadows every one of these RPOs, fleetingly but repeatedly, spending a substantial fraction of time in their neighborhoods.
We find that turbulence tends to shadow RPOs for a fraction of the period and that, when it does so, turbulence often shadows multiple ECSs at once. Given that turbulence never approaches any solution particularly closely, it is surprising how well and how often these solutions predict the evolution of turbulent flow.

With the sparse library of ECS used here, we are only able to capture a semi-quantitative description of turbulence. For instance, in the example shown in \autoref{fig:shadowing_tutorial}(a), both the energy and dissipation rate are found to differ rather substantially from those characterized by RPO$_1$, despite the flow fields appearing almost indistinguishable. However, as more ECS are considered, turbulence will make closer visits to individual solutions, and the accuracy with which observables are predicted during shadowing will increase. Interestingly, it is possible to predict turbulent averages somewhat accurately using a relatively small collection of shadowed orbits \cite{Yalniz2020}. 

Shadowing analysis opens new doors for describing turbulence in terms of a variety of invariant sets. The approach used here can be extended for detecting shadowing of other types of recurrent flows as well as non-recurrent flows. In particular, turbulence has been observed \cite{Suri2017, Suri2018} to shadow unstable manifolds as well.
Heteroclinic connections \cite{Suri2019,Farano2019}, which are the intersections of a stable and an unstable manifold that describe the dynamical transition from one ECS to another, are likely shadowed as well. Detecting heteroclinic shadowing events could strengthen the predictive power of Hopf's picture. Heteroclinic connections define the possible sequences of ECS that turbulence can shadow. Heteroclinic connections could also play an important role in describing extreme events, which are not recurrent and unlikely to be described by ECS. 

ECS have already generated much insight into fluid turbulence. 
For instance, ECS capture self sustaining processes that maintain wall-bounded turbulence \cite{Waleffe1997}) and elucidate the transition to turbulence \cite{itano2001,kreilos2012}. This study provides experimental validation that ECSs describe the spatiotemporal structure of sustained turbulence; basically, ECS serve as a backbone for turbulence \cite{cvitanovic2013}. 
The ability to detect shadowing is also crucial for implementing reduced descriptions of turbulence, such as symbolic dynamics \cite{Yalniz2020}.
Invariant sets such as equilibria, periodic orbits, heteroclinic connections, etc., collectively define a road map of turbulence.
Further development of a dynamical framework based on invariant sets \cite{Krygier2021, Crowley2022,Suri2019,Yalniz2020} could eventually enable forecasting and control of turbulence. 

\vskip6pt

\enlargethispage{20pt}


\dataccess{Insert details of how to access any supporting data here.}

\aucontribute{C.J.C. led experimental work, and contributed to the manuscript. J.L.P. led theoretical and numerical work, and
contributed to the manuscript. W.T. supported experimental and numerical work, and contributed to the manuscript. R.O.G. supervised theoretical and numerical work, and contributed to the manuscript. M.F.S. supervised experimental work, and contributed
to the manuscript. All authors read and approved the manuscript.}

\competing{The authors declare that they have no competing interests.}

\funding{We gratefully acknowledge financial support by ARO under grants W911NF-15-1-0471 and W911NF-16-10281 and by NSF under grant CMMI-1725587.}

\ack{Numerical simulations used the Taylor-Couette flow code written by Marc Avila. Michael C. Krygier developed numerical tools for finding and continuing recurrent solutions. }



\bibliographystyle{apsrev4-2}
\bibliography{references}

\end{document}